\documentclass[12pt]{iopart}
\usepackage{iopams}  
\usepackage{color}
\usepackage{tabularx}
\usepackage{epsfig}
\usepackage{graphicx}

\begin{document}
\title{Local field distributions in spin glasses}

\author{Stefan Boettcher}
\address{Physics Department, Emory University, Atlanta, Georgia
30322, USA}
\ead{sboettc@emory.edu}

\author{Helmut G.~Katzgraber} 
\address{Theoretische Physik, ETH Z\"urich, CH-8093 Z\"urich,
Switzerland}
\ead{katzgraber@phys.ethz.ch}

\author{David Sherrington} 
\address{Rudolf Peierls Centre for Theoretical Physics, University
of Oxford, 1 Keble Road, Oxford, OX1 3NP, UK}
\ead{d.sherrington1@physics.oxford.ac.uk}

\date{\today}

\begin{abstract} 
Numerical results for the local field distributions of a family
of Ising spin-glass models are presented. In particular,
the Edwards-Anderson model in dimensions two, three, and
four is considered, as well as spin glasses with long-range
power-law-modulated interactions that interpolate between a
nearest-neighbour Edwards-Anderson system in one dimension and the
infinite-range Sherrington-Kirkpatrick model. Remarkably, the local
field distributions only depend weakly on the range of the interactions
and the dimensionality, and show strong similarities except for near
zero local field.
\end{abstract} 

\maketitle 

\section{Introduction} 
\label{sec:introduction}

There has been interest in the distribution $P(h,T)$ of local fields
$h$ at temperature $T$ in spin glasses since the earliest days of
their theoretical study \cite{marshall:60,klein:63}. Particularly
influential was Thouless, Anderson and Palmer's \cite{thouless:77}
(TAP) self-consistent solution of $P(h,T = 0)$ for the
Sherrington-Kirkpatrick \cite{sherrington:75} (SK) infinite-ranged
spin-glass model for which a mean-field theory is believed to be
exact, albeit unusual and very subtle \cite{mezard:87,talagrand:03}.
Since then there have been several further studies of $P(h)$ for the
SK model (see, for example, Refs.~\cite{palmer:79}, \cite{thomsen:86},
\cite{oppermann:05}, and \cite{oppermann:07}) and the nature of the
local field distributions is well understood.  On the other hand,
there has been little work on the study of $P(h)$ for other spin-glass
models. Most notably, few studies exist of $P(h)$ for the finite-range
canonical Edwards-Anderson \cite{edwards:75,sherrington:75a} (EA)
Ising spin-glass model which is generally not exactly solvable. It
remains controversial whether some of the specific subtleties of the SK
model are applicable to the EA model and other more realistic models,
and their relationship is far from clear.

This paper studies and compares numerically the local field
distributions at $T=0$ of Ising spin glasses with varying
range-behaviour and spatial dimensionality, which are largely
inaccessible to exact solution.  We focus our discussion to Gaussian
bond distributions of zero mean.  A remarkable and somewhat surprising
similarity is found across systems for which other aspects of the
statistical physics state structure are believed to be different,
but with systematic small differences near $h=0$ as the system
interpolates between the limits of one-dimensional nearest neighbour
and infinite-ranged mean field, at both extremes of which $P(h=0,T=0)$
is zero in the thermodynamic limit.

The paper is structured as follows: In Sec.~\ref{sec:models} we
introduce the models studied followed by brief descriptions of the
numerical methods in Sec.~\ref{sec:numerics}. Our results are presented
in Sec.~\ref{sec:loc} followed by concluding remarks.

\section{Models}
\label{sec:models}
 
Both the SK model and the EA models are characterized by the Hamiltonian
\begin{equation}
{\mathcal H}= - \sum_{i<j} J_{ij} S_i S_j ,
\label{eq:SGmodel}
\end{equation} 
where $S_i \in \{\pm 1\}$. The interactions $J_{ij}$ are chosen
randomly and independently from a Gaussian distribution of zero
mean and then quenched. In the SK model the sum over $i$ and
$j$ extends over all sites and the variance of the distribution
${\mathcal P}(J_{ij})$ scales inversely with the system size $N$
as $J^{2}/N$. Here we set $J = 1$ so that the spin glass onset
transition is at $T_c = 1$. For the EA model the sum over the indices
$i$ and $j$ is restricted to nearest-neighbour pairs and the variance
of the bonds $J_{ij}$ is independent of the number of spins $N$
on a $d$-dimensional hyper-cubic lattice of size $N = L^d$; in this
case, unlike for the SK model, the lattice dimension is relevant
both qualitatively and quantitatively.  There is universal agreement
that the SK model exhibits a phase transition as the temperature is
reduced \cite{binder:86} to a phase with an ultrametric hierarchy
of metastable states and an associated mathematical feature of
replica-symmetry breaking of the overlap order parameter. For the
finite-range EA model there is a spin-glass transition above the
lower critical dimension, $d > d_l$, believed\cite{boettcher:05d}
to be $d_l=5/2$, but the existence of an ultrametric hierarchy in
short-range systems is controversial, not proven and disbelieved by
many practitioners.

\begin{figure}[!tbp]

\vskip 2.6in
\includegraphics{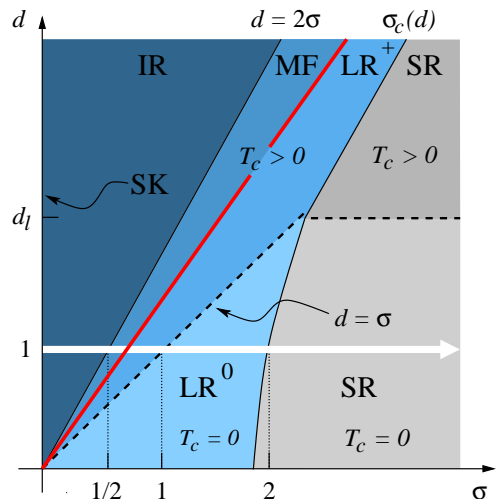}

\caption{ 
Schematic phase diagram of the KAS model in the $d$-$\sigma$ plane
following Ref.~\cite{fisher:88}. In this work we focus on $d = 1$
which corresponds to the white horizontal arrow. For $\sigma \le
1/2$ we expect infinite-range (IR) behaviour reminiscent of the SK
model and where the energy needs to be rescaled as a power of the
system size to avoid divergences, whereas for $1/2 < \sigma \le 2/3$
we have mean-field (MF) behaviour corresponding to an effective space
dimension $d_{\rm eff} \ge 6$ (the thickened line separates mean-field
from non-mean-field behaviour). For $2/3 < \sigma \lesssim 1$ we have
a long-range (LR$^+$) spin glass with a finite ordering temperature
$T_{\rm c}$, whereas for $1 \lesssim \sigma \lesssim 2$ we have a
long-range spin glass with $T_{\rm c} = 0$ (LR$^0$). For $\sigma
\gtrsim 2$ the model displays short-range (SR) behaviour with a zero
transition temperature \cite{kotliar:83,katzgraber:03}. Empirically,
for $0.5 \le \sigma \lesssim 1$ the $d=1$ KAS model can be identified
as corresponding to an EA system with effective dimension $d_{\rm
eff} \approx 2/(2\sigma - 1)$ \cite{binder:86}. Figure adapted from
Ref.~\cite{katzgraber:03}.}
\label{fig:dsigma}
\end{figure}

In order to effectively interpolate between the physics
of the SK and EA models and to probe their similarities and
differences we also study a ``tunable'' model first introduced
by Kotliar, Anderson and Stein (KAS)~\cite{kotliar:83}
and recently studied in detail by Katzgraber and Young
\cite{katzgraber:03,katzgraber:03f,katzgraber:04c,katzgraber:05c}.  The
model, which has helped elucidate many properties of spin glasses, is
a long-range Ising spin glass with random power-law interactions. The
Hamiltonian of the model is given by Eq.~(\ref{eq:SGmodel}) but now
with the sites $i$ and $j$ on a $d$-dimensional lattice with periodic
boundary conditions and the exchange interactions given by
\begin{equation}
J_{ij}= c({\sigma}) \frac{\epsilon_{ij}}{{{r_{ij}}^\sigma}}.
\label{eq:JKY}
\end{equation} 
Here, $r_{ij}$ is the separation of spins $i$ and $j$, the
$\epsilon_{ij}$ are chosen randomly and independently from a
Gaussian distribution of zero mean and standard deviation unity, and
$c(\sigma)$ is a constant. The KAS model is believed to interpolate
between mean-field-like behaviour for small $\sigma < \sigma_{\rm
c1}(d)$, an intermediate long-range regime [$\sigma_{\rm c1}(d)
<\sigma <\sigma_{\rm c2}(d)$], and a short-range regime [$\sigma_{\rm
c2}(d)<\sigma < \sigma_{\rm c3}(d)=\infty$].  Each of the latter two
regimes are subdivided into ordering and non-ordering regimes depending
upon the space dimension (higher dimensions favoring ordering); see
Fig.~\ref{fig:dsigma}.  In the present work we shall consider only
$d = 1$ for the KAS model, for which case the intermediate long-range
region has a finite cooperative ordering temperature $T_{\rm c} > 0$
for $\sigma < \sigma_{\rm c}$ but no finite-temperature ordering for
$\sigma > \sigma_{\rm c}$; see Fig.~\ref{fig:dsigma}. To enforce
periodic boundary conditions we place the spins on a ring (see
Ref.~\cite{katzgraber:03} for details) and choose the geometric
distance between the spins, i.e., $r_{ij} = (N/\pi)\sin(\pi |i -
j|/N)$.  We normalize the interactions to have $T_{\rm c}^{\rm MF}
= 1$ for all $\sigma$, i.e.,
\begin{equation} 
c({\sigma})^{-2} = \sum_{j\neq i}r{_{ij}}^{-2\sigma}.
\label{eq:cnorm}
\end{equation}

\begin{figure}[!tbp]
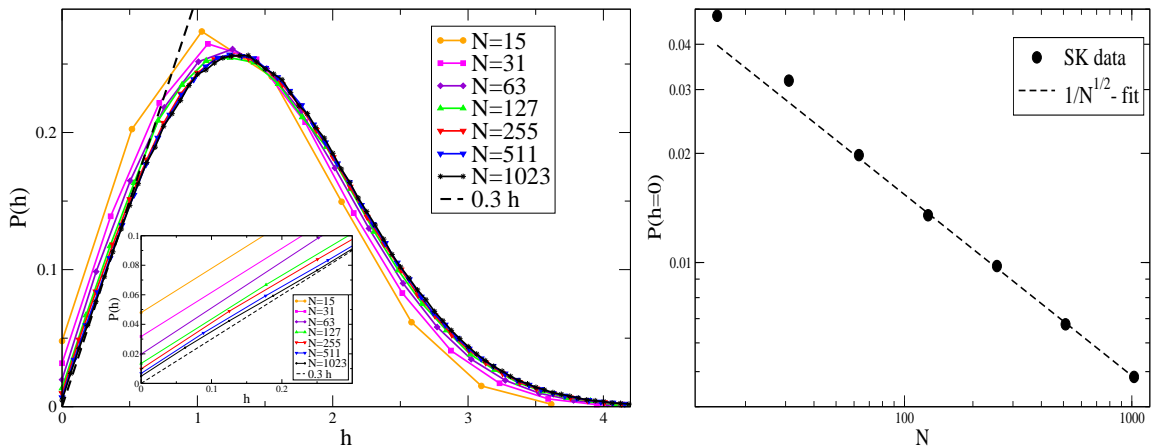

\vskip 2.4in
\includegraphics{Ph.eps}
\includegraphics{Ph0.eps}
\includegraphics{Ph0_extra.eps}

\caption{
Left panel: $P(h)$ for $h\geq0$ for SK systems of sizes
$N=15,\ldots,1023$ at $T=0$. (Here, a bimodal bond distribution,
$J=\pm1/\sqrt{N}$, has been used.) Clearly, the curves are well
converged for the larger system sizes. The inset shows an enlargement
of $P(h)$ for $h$ near zero. The slopes are all $\approx 0.3$, as
indicated by the dashed line. Right panel: $P(h = 0) \sim 0.153(3)
N^{-1/2}$, i.e., the data extrapolate to zero in the thermodynamic
limit.  The low noise in the data suggest that errors are smaller than the symbols.
}
\label{fig:PhSK}
\end{figure}

For the short-range EA model we study hyper-cubic lattices in
different space dimensions $d$.  Again, in order to assist quantitative
comparisons, in all cases we choose the exchange-scale normalization
to yield the same mean-field transition temperature $T_{\rm c}^{\rm
MF}=1$. Hence for EA models we choose the variance of the exchange to
be $1/z$ where $z$ is the coordination number; for the hyper-cubic
lattices that we study $z=2d$. In the thermodynamic limit the SK
model is equivalent to the infinite-dimensional EA model; hence its
normalization with the variance of the exchange interactions scaling
as $1/N$.

\section{Numerical procedures}
\label{sec:numerics}

For the EA model we apply the Extremal Optimization (EO) heuristic
as described in Ref.~\cite{boettcher:01}. EO provides approximate
ground states of spin glasses with high accuracy typically within
$O(N^3)$ update steps, at least for system sizes up to $N\leq256$
as studied here for the EA model. We read off and average the local
fields for the presumed zero-temperature configuration found for each
instance. For each reported system, between $10^4-10^5$ instances have
been optimized, depending on system size. Since EO finds near-optima
using a far-from-equilibrium dynamics, the explored configurations
may possess a systematic bias. To check that this is not the case we have
applied
EO to reproduce $P(h)$
for the SK model (with $J_{ij} \in\{\pm N^{-1/2}\}$) for $N\leq1023$,
as studied in Ref.~\cite{boettcher:05}, finding no such biases.
This is apparent from Fig.~\ref{fig:PhSK}, in which familiar properties
(discussed below) of $P(h)$ in the SK model are reproduced, such
as the linear slope for $|h|\to0$ and the finite-size scaling of
$P(h=0)\sim1/\sqrt{N}$ for $N\to\infty$.

For the KAS model we use exchange (parallel tempering) Monte Carlo
for the simulations \cite{hukushima:96}. To measure the local field
distributions we compute 5000 disorder realizations for each system
size $N$ and value of $\sigma$. The lowest temperature simulated
is $T = 0.05$ which is close enough to $T = 0$ such that for the
system sizes studied we effectively probe the ground state of
the system \cite{comment:gs}. We also note that earlier studies
of the SK model \cite{thomsen:86} have shown that $P(h = 0,T) =
{\lambda}T + O(T^2)$ with $\lambda \approx (2 \pi e)^{-1/2} \approx 0.25$, 
giving a deviation
of $\approx0.01$ for the SK model at $T=0.05$ compared with $T=0$,
which is indeed negligible compared to the values we find for small
$h$ for finite-range models.

In this study we consider $\sigma = 0.00$ (SK), $0.55$ (MF), $0.75$
(LR$^+$), $0.83$ (LR$^+$), $1.00$ as well as $1.50$ (LR$^0$) and
$2.00$ (SR); see Fig.~\ref{fig:dsigma} for details. For all system
sizes $N$ and exponents $\sigma$ we study $N_T = 29$ replicas. For
$\sigma \le 1.0$  we equilibrate the system for $N_{\rm sw} =
2^{17}$ Monte Carlo sweeps for $N \le 128$ and for $2^{20}$ Monte
Carlo sweeps for $N = 256$.  For $1.50 \le \sigma \le 2.0$ we again
take $2^{17}$ Monte Carlo sweeps to equilibrate for $N \le 64$ but
increase to $2^{19}$ Monte Carlo sweeps for $N = 128$.  Then in each
case we measure the local field distributions for the same amount
of Monte Carlo time. Equilibration is tested by comparing the energy
calculated from the link overlap to the internal energy of the system
calculated directly. Once both agree, the system is considered to be
in thermal equilibrium. For details see Refs.~\cite{katzgraber:05c}
and \cite{katzgraber:01}.

\section{Local field distributions}
\label{sec:loc}

The local field distribution is defined by
\begin{equation}
P(h) = \left[\left<  \frac{1}{N}
\sum_{i}\delta{\left(h-\sum_{j}J_{ij}S_{j}\right)} \right>\right]_{\rm av} ,
\label{eg:ph}
\end{equation}
where $\langle \cdots \rangle$ indicates a thermodynamic average and
$[\cdots]_{\rm av}$ denotes an average over the quenched disorder.

The simplest nontrivial mean-field solution for
$P(h)$ for the SK or EA models, the replica-symmetric
\cite{edwards:75,sherrington:75,schowalter:79} effective field
approximation, yields $P(h)$ self-consistently through
\begin{eqnarray}
P(h)= \frac{1}{\sqrt{2\pi q}}e^{-{h^2}/2q},\qquad
q=\int {\rm d}h P(h) \tanh^2({\beta}h) ,
\label{eq: P_H_RS}
\end{eqnarray}
where $\beta = 1/T$ is the inverse temperature. However, this
approximation incorrectly yields a hole in $P(h)$ as $T \rightarrow
0$ \cite{schowalter:79}, i.e., $P(h)=0$ for all $\mid h \mid  <
(2/\pi)^{1/2}$.

Indeed, Thouless, Anderson and Palmer \cite{thouless:77} argued
already in 1977 that for small fields $h$ in the SK model $P(h,T
= 0) \sim 0.3|h|$, linearly in $|h|$ \cite{comment:bound}. Later
studies of the SK model, employing full replica symmetry breaking
(FRSB) \cite{parisi:80a}, have borne out this linear form and
the value of its slope (see, for example, Fig.~\ref{fig:PhSK}
and Refs.~\cite{thomsen:86}, \cite{oppermann:07}, and
\cite{pankov:06}). Through accurate studies of a large sequence
of finite-replica-symmetry breakings Refs.~\cite{oppermann:05} and
\cite{oppermann:07} have demonstrated that within replica theory the
correct linear behaviour of $P(h)$ for the SK model requires the full
limit of infinite replica-symmetry breaking order; any finite-order
truncation yielding a fictitious (if decreasing with RSB-order)
hole in $P(h)$ near $h = 0$.

Another model that is exactly solvable is the one-dimensional
nearest neighbour random-exchange Ising chain (limit of the
KAS model for $\sigma \gg 1$), the one-dimensional EA model
\cite{barma:79,thomsen:86}.  It does not, however, have either
frustration or a finite-temperature phase transition and, thus,
no spin glass phase. $P(h)$ is given by
\begin{equation} 
P(h) = \int {\rm d}J  {\mathcal P}(J){\mathcal P}(h+J) 
 \left\{1-\tanh[\beta J]\tanh\left[\beta(h+J)\right]\right\},
\end{equation}
with 
\begin{equation}
{\mathcal P}(J) = \frac{1}{\sqrt{\pi}}e^{-J^2},
\end{equation} 
such that $\langle J^2\rangle=1/2$. For $\beta\to\infty$, this 
evaluates to
\begin{equation}
P(h) = \frac{2}{\pi}|h|\int_0^{1}{\rm d}x
       e^{-h^2\left(2x^2-2x+1\right)},
\label{1dexacteq}
\end{equation} 
again giving $P(h = 0) = 0$ and a linear small-$h$ behaviour,
analogously to the SK model, but with a slope almost twice as large,
i.e., $P(h) \sim 2|h|/\pi$.

\begin{figure}[!tbp]
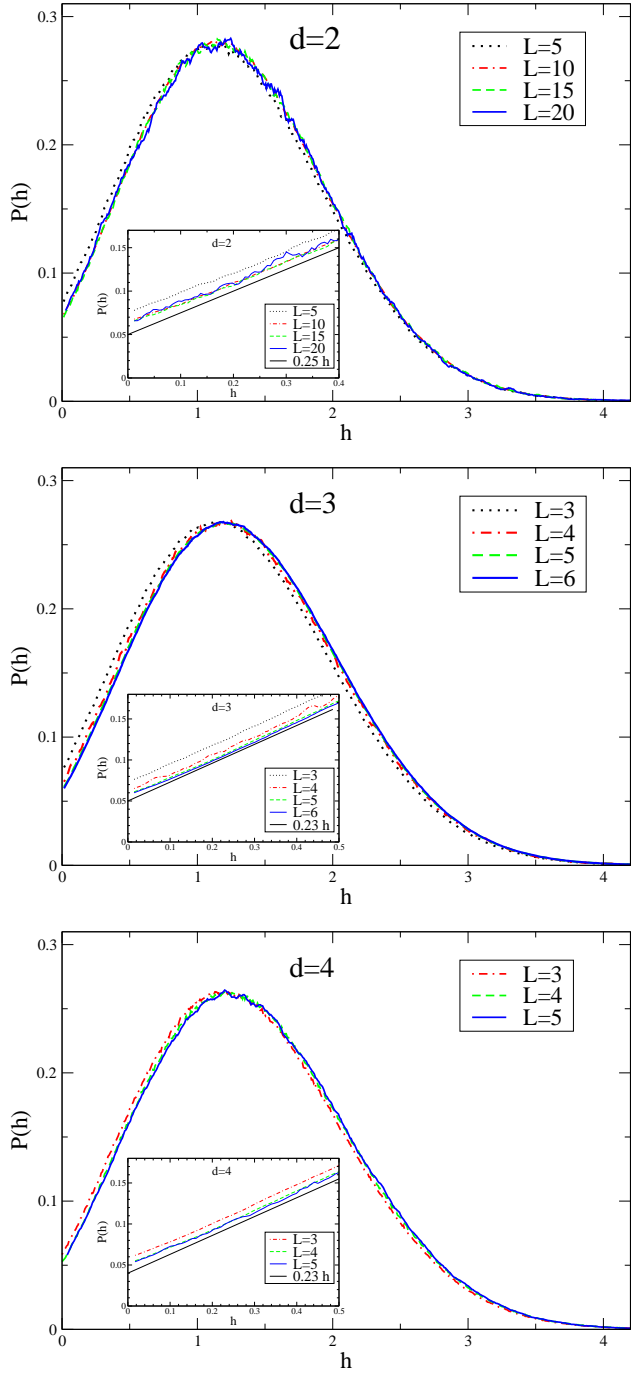


\vskip 7.2in
\includegraphics{Ph_D2.eps}
\includegraphics{Ph_D2h0.eps}
\includegraphics{Ph_D3.eps}
\includegraphics{Ph_D3h0.eps}
\includegraphics{Ph_D4.eps}
\includegraphics{Ph_D4h0.eps}

\caption{
$P(h)$ for $h>0$ for the EA spin glass on hyper-cubic lattices for
$d=2$, $3$, and $4$ for various side-lengths $L$ for a Gaussian
bond distribution with $\langle J^2\rangle=1/(2d)$. Already beyond
some small $L$, there is little distinction between the $P(h)$ for 
different sizes, 
indicating only small corrections to scaling. This
becomes even more apparent in the insets, showing an enlargement
of the data near $h=0$. $P(h)$ in each $d$ seems to converge to a
finite value at $h=0$ of $P(h=0) \approx 0.065$ ($d=2$), $P(h=0)
\approx 0.055$ ($d=3$), and $P(h=0)\approx 0.045$ ($d=4$), see
Fig.~\protect\ref{fig:extraPh0}. $P(h)$ for small $|h|$ rises with
a slope of $a \approx0.25$ ($d=2$), $a \approx0.23$ ($d=3$), and $a
\approx0.23$ ($d=4$), where $P(h) \sim P(0) + a|h|$.
}
\label{fig:Phd}
\end{figure}

\begin{figure}[!tbp]

\vskip 2.6in
\includegraphics{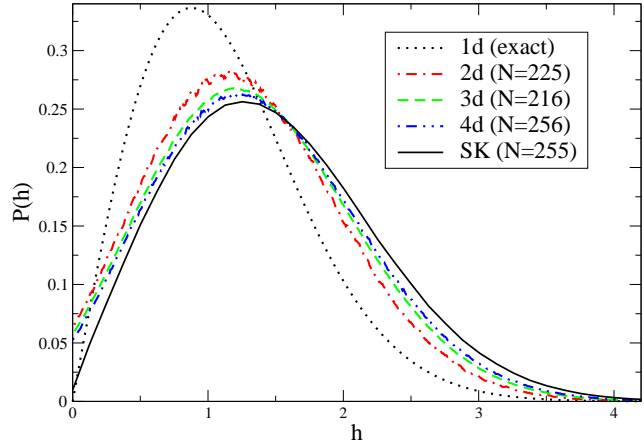}

\caption{
$P(h)$ for $h>0$ for the EA spin glass in $d = 2$, $3$, and $4$
dimensions and for the SK model, all at comparable system size
$N=216\ldots256$ for bond distributions with $\langle J^2\rangle=1/z$,
where $z$ is the connectivity of each spin: $z=2d$ for the EA model and
$z=N-1$ for the SK model.  Also plotted is the {\it exact} $d=1$ result
from Eq.~(\protect\ref{1dexacteq}).  The plot highlights the overall
similarity in $P(h)$ for all models and space dimensions. Except
for $|h|$ close to zero, the numerical data for the EA model in $d=2$ -- $4$
overall seem to interpolate monotonically  between the $d=1$ result
and the SK ($d=\infty$) model. Yet, while $P(0)=0$ for both $d=1$
and SK, for $d=2$ -- $4$, $P(0)$ is positive and monotonically decreasing,
see Fig.~\protect\ref{fig:extraPh0}.
}
\label{fig:PhAll}
\end{figure}
\begin{figure}[!tbp]

\vskip 2.6in
\includegraphics{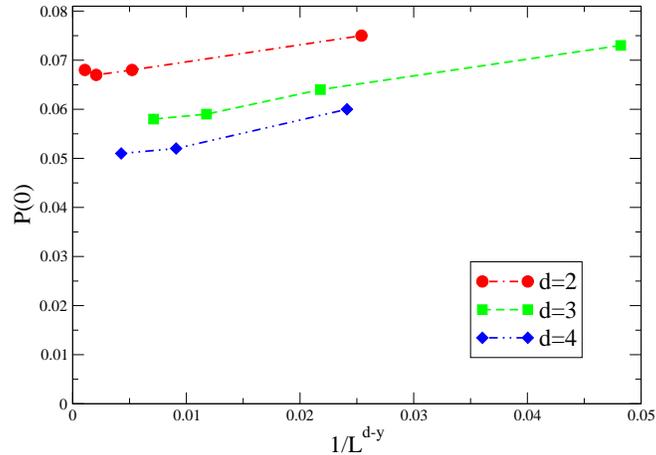}

\caption{
Extrapolation of the values for $P(0)$ obtained for $d=2$, $3$,
and $4$ in Figs.~\protect\ref{fig:Phd} for
various system sizes $L$. The extrapolation to the thermodynamic
limit $L\to\infty$ proceeds on a scale of $1/L^{d-y}$, where
$y$ for each $d$ is the stiffness exponents discussed in
Ref.~\protect\cite{boettcher:05d}. As $P(h)$ has units of
inverse energy, these are the appropriate corrections to scaling for
the EA model \cite{boettcher:08,campbell:04}. Clearly, $P(0)>0$ for
each $d$ at $L\to\infty$, unlike for the SK model. The noise in the
data suggests
errors to be about double the size of the symbols.
}
\label{fig:extraPh0}
\end{figure}

\begin{figure*}[!tbp]
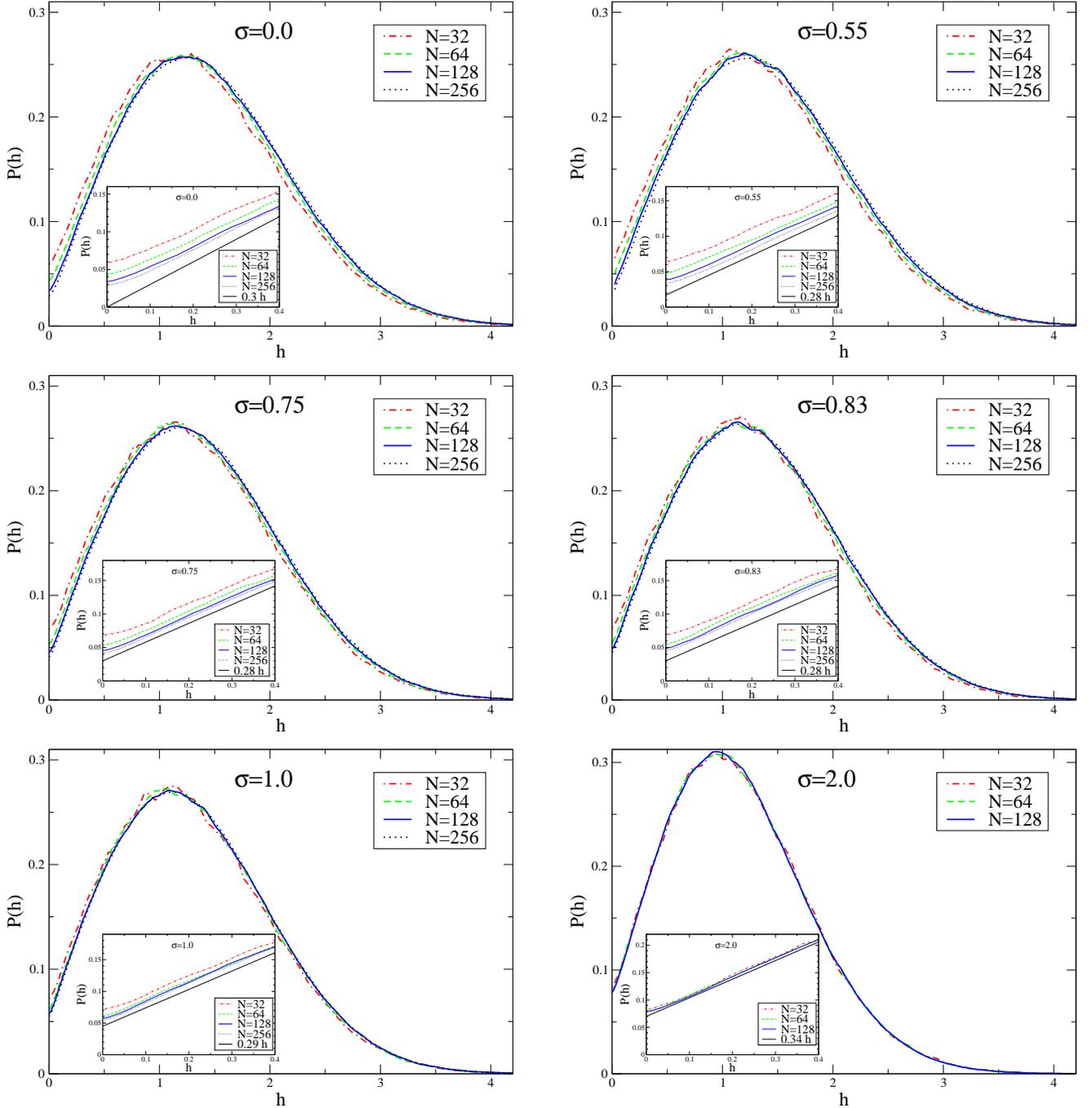


\vskip 2.4in
\includegraphics{Ph_sigma000.eps}
\includegraphics{Ph_sigma000h0.eps}

\includegraphics{Ph_sigma055.eps}
\includegraphics{Ph_sigma055h0.eps}

\vskip 2.4in
\includegraphics{Ph_sigma075.eps}
\includegraphics{Ph_sigma075h0.eps}

\includegraphics{Ph_sigma083.eps}
\includegraphics{Ph_sigma083h0.eps}

\vskip 2.4in
\includegraphics{Ph_sigma100.eps}
\includegraphics{Ph_sigma100h0.eps}

\includegraphics{Ph_sigma200.eps}
\includegraphics{Ph_sigma200h0.eps}

\caption{
$P(h)$ for $h>0$ and different system sizes $N$ for the KAS model in $d
= 1$ for different powers of the exponent $\sigma$ which change the
effective space dimension. The data are for the  lowest temperature
simulated, $ T = 0.05$. The insets show an enlargement of the area
around $h = 0$. $P(h = 0)$ clearly converges to a finite value for
all $\sigma > 0.5$ and shows an approximately linear behaviour $[P(h)
- P(0)]\sim 0.3 |h|$ for small $|h|$. Note that the corrections to
scaling decrease considerably for larger $\sigma$ values.
}
\label{fig:Phsigma}
\end{figure*}

In higher dimensions the EA model is not exactly solvable. There is
no finite-temperature spin-glass phase in $d=2$, although there is
such a phase \cite{katzgraber:06} in $d=3$ and greater. Yet, it is not
clear that the glassy phase for $d\geq3$ exhibits the characteristics
of replica-symmetry breaking (e.g., ultrametricity) found in the SK
model; at least up to an upper critical dimension believed 
to be $d_{\rm ucd}=6$. We have performed a numerical simulation of $P(h)$ for the
EA model using the EO heuristic \cite{boettcher:01}. The results are
exhibited in Figs.~\ref{fig:Phd}. [Since $P(h)$ is symmetric, we only
show plots for $h\geq0$, i.~e., $\int_0^\infty P(h){\rm d}h=1/2$.]
At first sight $P(h)$ at $T=0$ shows very little variation between the
dimensions $d$; the overall shape is quite similar to that for the
SK model, see Fig.~\ref{fig:PhAll}. But, in fact, at closer detail
there is a notable distinction for $h\to0$, where the behaviour
for the EA model differs significantly from the SK model. Unlike
for the SK model (see Fig.~\ref{fig:PhSK}), $P(h = 0)$ appears to
be finite in the thermodynamic limit of the EA model, as the plot
near $h=0$ in the bottom panels of Figs.~\ref{fig:Phd} indicate. We
have extrapolated the values of $P(0)$ to infinite system size $L$ in
Fig.~\ref{fig:extraPh0}.  For each $d\geq2$, whether above or below the
lower critical dimension, $P(0)$ quickly settles to a positive value
between approximately $0.04$ and $0.07$. This value seems to decrease
slowly with increasing space dimension $d$, consistent with $P(0)=0$
in the SK ($d=\infty$) limit; see Fig.~\ref{fig:extraPh0d} below.
$P(h)-P(0)$ rises linearly for small $h$ with a weakly $d$-dependent
slope of $a\approx 0.23$ to $0.25$.

\begin{figure}[!tbp]

\vskip 2.6in
\includegraphics{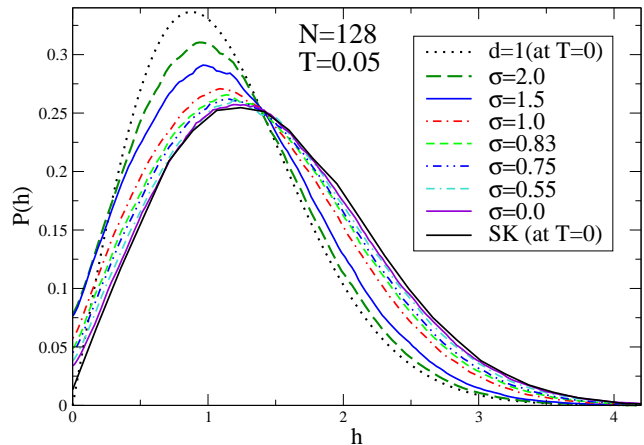}

\caption{
Direct comparison of the local field distribution $P(h)$ (at $T=0.05$)
as a function of the local field $h$ for the KAS model and $N = 128$
spins and different values of the exponent $\sigma$ covering all
possible universality classes from infinite range to short range with
zero transition temperature.  Also plotted are the exact results for
$d=1$ (i.e., $\sigma\to\infty$) from Eq.~(\protect\ref{1dexacteq})
and the SK data (i.e., $\sigma=0$) from Fig.~\protect\ref{fig:PhSK}
(both at $T=0$). As in Fig.~\protect\ref{fig:PhAll}, the data
interpolate smoothly between both extremes, except at $P(0)$.
}
\label{fig:KY}
\end{figure}

We have performed a similar set of simulations for the $d = 1$ KAS
model for a range of different $\sigma$ values covering different
behaviours ranging from SK-like to finite-range non-ordering
($T_{\rm c} = 0$) \cite{katzgraber:03}. In Fig.~\ref{fig:Phsigma}
we show the local field distributions for $T = 0.05$ for different
system sizes $N$.  Each panel is for a different exponent $\sigma$
in Eq.~(\ref{eq:JKY}). The similarity of all data sets for different
$\sigma$ is clearly visible. The insets show always the area around
$h = 0$ in detail. To further illustrate the similarities between
the data sets, in Fig.~\ref{fig:KY} we show data for $P(h)$ for
$N = 128$ and different exponents $\sigma$ covering all possible
universality classes.  The data for all $\sigma$ agree relatively
well, with the data for $\sigma > 1.0$ showing a more pronounced peak
and larger gap.  Again, for the different values of $\sigma > 0.5$
studied, $[P(h) - P(0)] \sim a|h|$ for $|h|$ small with $a \approx
0.3$ interpolating between the SK and the short-range one-dimensional
result \cite{comment:ab}.

\begin{figure}
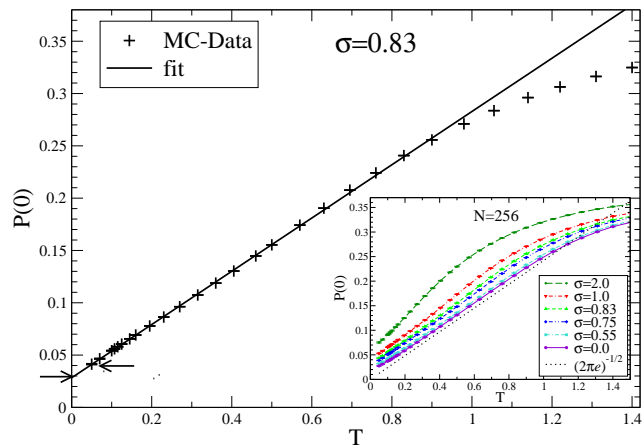


\vskip 2.6in
\includegraphics{TextraPh0_083.eps}
\includegraphics{P0L256allTallSigma.eps}

\caption{
$P(h=0)$ as a function of temperature for $\sigma = 0.83$ and $N = 256$.
The data are well fit by a linear behaviour in $T$ (with very small
quadratic corrections) with slope $\approx0.25$ for $T \le 0.3$. Data
for $T = 0.05$ are very close to $T = 0$ (arrows), which is why in
Fig.~\ref{fig:extraPh0_KY} we extrapolate to $N = \infty$ for $T =
0.05$ and not $T = 0$. Furthermore, the estimate of the error bars
in the temperature extrapolation is difficult. Inset: $P(h=0)$
for a range of $\sigma$ compared with the SK result $(2 \pi e)^{-1/2} T$.
}
\label{fig:TextraPh0_083}
\end{figure}

\begin{figure}

\vskip 2.6in
\includegraphics{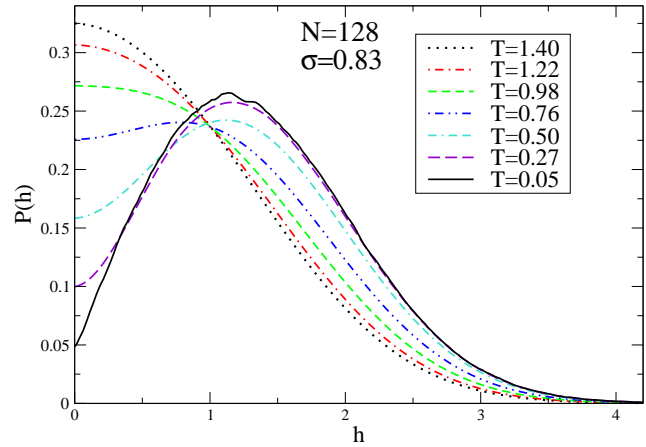}

\caption{
$P(h)$ for $\sigma = 0.83$ and $N = 128$ for different temperatures ranging from
$T = 0.05$ to temperatures well above the transition temperature
$T_{\rm c}$. For low enough $T$ the local-field distribution shows a
dip around $h = 0$ which is ``filled in'' for increasing temperatures.
For $T \gg T_{\rm c}$ we obtain a Gaussian distribution.
}
\label{fig:PhT_128_083}
\end{figure}

\begin{figure}[t]

\vskip 2.6in
\includegraphics{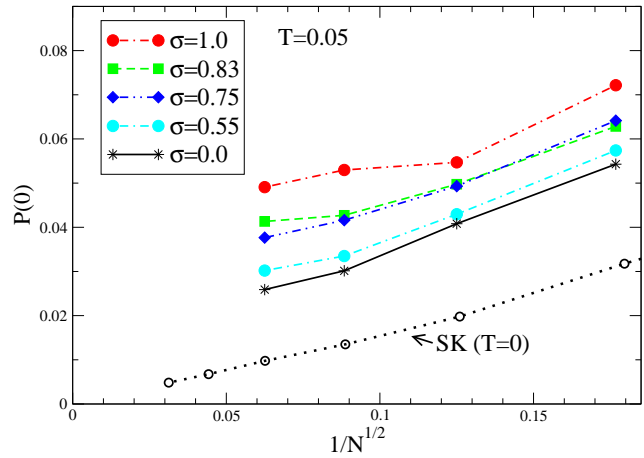}

\caption{
Extrapolation of the values for $P(0)$ obtained for the KAS model as
a function of $1/\sqrt{N}$ for $T = 0.05$ and different values of the
exponent $\sigma$. The data for $\sigma = 0.00$ decay with a power-law
similar to the SK model. For larger values of
$\sigma$ the data indicate a saturation to a finite value of $P(0)$
for $N \rightarrow \infty$. Larger system sizes would be needed to
clearly differentiate the different scenarios beyond a qualitative
comparison. The noise in the data suggests an error of about 5 times the
symbol size. For comparison we also show the data for SK at $T=0$
from Fig.~\protect\ref{fig:PhSK}; note that the shift compared with the 
$\sigma=0$ KAS results is due to the finite-temperature 
shift $(2 \pi e)^{-1/2}T$.
}
\label{fig:extraPh0_KY}
\end{figure}

We have also extrapolated the data for $P(h = 0,T,N)$ to $T =
0$ (fits to a quadratic function for $T \le 0.3$ with fitting
probabilities \cite{press:95} larger than $\sim 0.3$). A typical
example of the extrapolation for $\sigma = 0.83$ is shown in
Fig.~\ref{fig:TextraPh0_083} (the behaviour of $P(h)$ for different
temperatures is shown in Fig.~\ref{fig:PhT_128_083}).  Since the
difference between $T = 0.05$ and $T = 0$ is minimal and because
estimating the error bars for the extrapolated data is difficult, in
Fig.~\ref{fig:extraPh0_KY} we show data for $T = 0.05$.  
 
\begin{figure}
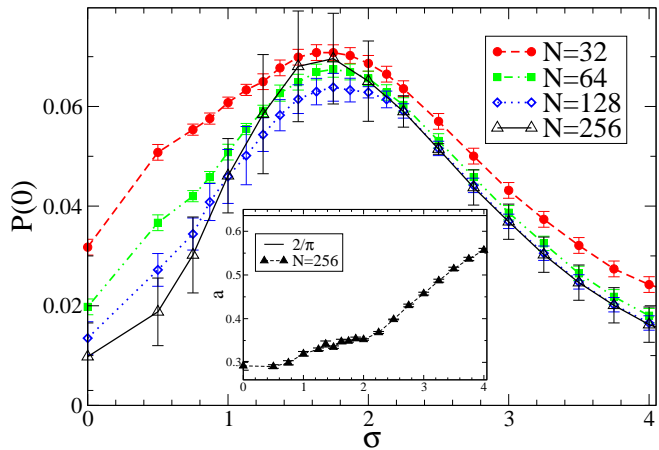


\vskip 2.6in
\includegraphics{sigma_max.eps}
\includegraphics{P0sigma_slope.eps}

\caption{
Plot of $P(0)$ as a function of $\sigma$ in the KAS model. For each
data point in $N=32$ and 64 averaged $P(h)$ over at least $10^5$
instances, $2\,10^4$ instances at $N=128$, and at least $10^3$
for $N=256$. (The data for $N=256$ clearly show a systematic bias
{\it only} near the maximum, indicative that those instances are
hardest to optimize. Note the complete lack of residual finite size
effects for $\sigma>2.0$). To obtain smooth estimates of $P(0)$,
independent of the width of histograms chosen to measure $P(h)$,
we made a linear fit to the respective $P(h)$ for $h<0.3$ to obtain
each data point. Inset: Plot of the slope $a = \partial P(h)/\partial
h|_{h\to0}$ of $P(h)$ near $h=0$ as a function of $\sigma$, extracted
from a fit to the same data. Only data for $N=256$ are shown, as there
are hardly any finite size effects. $2/\pi$ (solid line) represents
the slope for the one-dimensional result in Eq.~(\ref{1dexacteq})
obtained for $\sigma\to\infty$.
}
\label{fig:sigma_max}
\end{figure}

A closer look at Figs.~\ref{fig:KY} and~\ref{fig:extraPh0_KY} poses
an interesting question: In the KAS model, $P(0)$ appears to be
an increasing function of $\sigma$ for all values studied so far,
rising from $P(0)=0$ at $\sigma=0$ for SK to $P(0)\approx0.08$
at $\sigma=2.0$. Yet, the exact result for the one-dimensional
Ising chain, corresponding to $\sigma=\infty$, again has  $P(0)=0$.
Hence either the apparent extrapolation to the limit $N\to\infty$
limit, shown in Fig.~\ref{fig:extraPh0_KY}, is incorrect and
$P(0)=0$ for all $\sigma$ after all, or there has to be a maximum
in $P(0)$ at some finite $\sigma_{\rm max}$ beyond which $P(0)$
again descends to zero. To decide this question, we have done
a more extended study of $P(0)$ also for $\sigma>2.0$ which is
shown in Fig.~\ref{fig:sigma_max}. The results clearly show a
well-defined maximum near $\sigma_{\rm max}\approx1.8$, which
is likely to persist in the thermodynamic limit. The resolution
of this question---surprising in its own right---strengthens our
belief that finite-size effects in our presentation are well under
control. In all, the KAS model essentially reproduces the results
found for the SK and canonical EA models for $\sigma \lesssim 1$
where it corresponds to possible physical dimensions and for larger
$\sigma$ [where the formula $d_{\rm eff} \approx 2/{(2\sigma -1)}$
\cite{binder:86} becomes inappropriate] it continues smoothly towards
the nearest-neighbour $d=1$ EA limit. The fact that $P(0)$ peaks
for the KAS model at a value of $\sigma$  intermediate between that
corresponding to $d_{\rm eff} = 2$ and $d_{\rm eff} = 1$ suggests
that if one could continue the EA model off integer dimensions, there
might exist a dimension $d_{\rm max}$, likely between $d=1$ and $d=2$
at which $P(0)$ would peak. The slope of $[P(h)-P(0)]$ near $h=0$
is however found to vary monotonically with $\sigma$ between the SK
($\sigma =0$) and nearest neighbour EA ($\sigma = \infty$) limits.

The local field distribution $P(h)$ for a disordered Ising spin system
with Gaussian distributed random exchange disorder is thus seen to
be remarkably robustly linear in $|h|$ at low $h$ with a coefficient
in the range $a=0.25$ -- $0.35$, irrespectively of whether the system
exhibits spin-glass order and independently of whether such spin-glass
order is replica-symmetry-broken mean field or not.

It is also interesting to compare with  studies of  a non-equilibrated
SK model \cite{parisi:95,eastham:06,horner:07}.  In these studies,
spins are initially randomized and then exposed to purely relaxational
single-spin dynamics, i.e., dynamics in which spins are chosen randomly
and flipped if and only if such a flip would reduce the energy, until
a metastable state is reached. The distribution $P(h)$ averaged over
only those metastable states reached by this procedure is again linear
in $|h|$ for small $h$ with a slope that again appears to be of order
$a\sim0.3$.  The interest of this observation in the present context
is that the metastable states reached in this dynamical procedure
are not the ground state. Indeed they are significantly above the
ground state, with the average relaxational energy variously 
reported as  $E_{\rm relax}/N
\approx -0.7$ \cite{eastham:06}, -0.715 \cite{parisi:95}, and $-0.73$
\cite{horner:07}, much higher than the ground state energy per spin
$E_{\rm gs}/N = -0.7632$ \cite{oppermann:07}, while the corresponding
$P(h, E_{\rm relax})$ averaged  over all metastable states has a
finite value at $h=0$ \cite{parisi:95,eastham:06}.

\begin{figure}
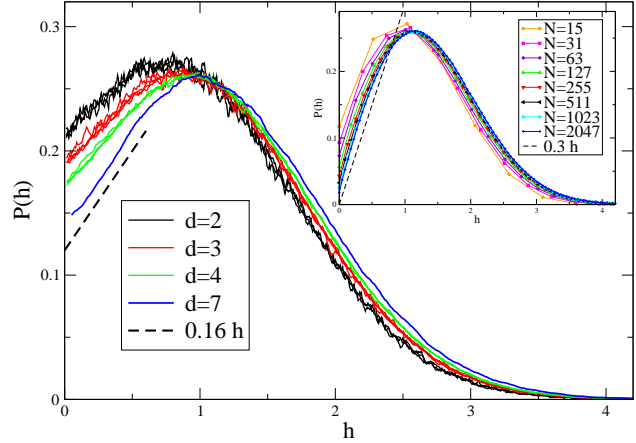


\vskip 2.6in
\includegraphics{PhEA_Ballistic.eps}
\includegraphics{PhSK_ballistic.eps}

\caption{
Local field distribution $P(h)$ for the EA model in $d=2$, $3$, and
$4$, obtained after purely relaxational single-spin dynamics on $10^4$
instances for each $L$ and $d$. The same system sizes $L$ as in
Figs.~\protect\ref{fig:Phd} have been used here but finite-size
effects are well below the statistical noise. We have added results
for $d=7$ ($L=3$) to show that there are no drastic changes above the
upper critical dimension $d_{\rm ucd}=6$. The dashed line merely serves to
guide the eye. Inset: $P(h)$ for relaxational single-spin dynamics in SK.
}
\label{fig:PhEA_Ballistic}
\end{figure}

\begin{figure}

\vskip 2.6in
\includegraphics{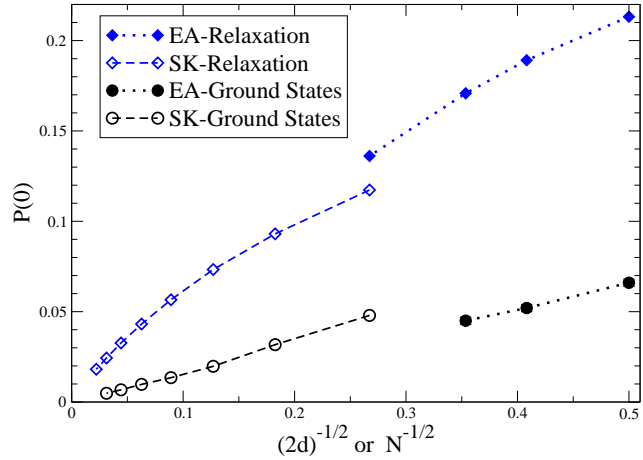}

\caption{
Extrapolation of $P(0)$ towards the SK-limit, $d\to\infty$. Shown
are the values of $P(0)$ obtained from the $L\to\infty$ extrapolation
in Fig.~\protect\ref{fig:extraPh0} for ground states of the EA model
(full circles) and those obtained by a relaxation process of the data
shown in Fig.~\protect\ref{fig:PhEA_Ballistic} (full diamonds), which
are approximately $L$-independent. Added are the data points for SK
for ground states from Fig.~\protect\ref{fig:PhSK} (open circles)
and for relaxation from Fig.~\protect\ref{fig:PhEA_Ballistic}
(open diamonds). As the EA spin glass approaches the SK model
by setting $2d=N-1\to\infty$, the appropriate scale here is
$(2d)^{-1/2}\sim N^{-1/2}$, according to the right panel of
Fig.~\protect\ref{fig:PhSK}.  Asymptotic slopes for the SK data are
$0.153(3)$ for ground states and approximately $0.61$ for relaxation
(see also Ref.~\protect\cite{horner:07}). For both sets of data,
it seems that ${\lim_{d \to\infty}}P(0)\to0$, consistent with
the respective SK results. Similar to the case of $\sigma>0.5$ in
Fig.~\protect\ref{fig:sigma_max}, there is no indication here that
$P(0)=0$ for any $d<\infty$.
}
\label{fig:extraPh0d}
\end{figure}

We have conducted a comparable study of $P(h)$ over metastable
states reached by rapid-quench for the EA model in $d=2$, $3$, $4$,
and $7$, and SK, as shown in Fig.~\ref{fig:PhEA_Ballistic}. As for
the results for the SK model, there are significant (qualitative)
similarities between the $T=0$ equilibrium results of $P(h)$ in
Figs.~\ref{fig:Phd} and those obtained by simple relaxation. But
especially the behaviour near $h=0$ deviates quantitatively,
with $P(0)$ distinctly larger and the initial slope significantly
smaller. Yet, both seem to tend smoothly towards the corresponding SK
result $P(0)=0$ (see Fig.~\ref{fig:PhSK} and Refs.~\cite{parisi:95}
and \cite{eastham:06}, respectively) for $d\to\infty$, as is
demonstrated in Fig.~\ref{fig:extraPh0d}.  Furthermore, for small
$|h|$ the local field distribution shows a linear behaviour as found
in Ref.~\cite{pazmandi:99} via an intrinsically far-from-equilibrium
simulation along the hysteresis loop of the model for finite external
fields.  Finally, as in Refs.~\cite{parisi:95} and \cite{eastham:06},
the metastable states obtained under relaxation in the EA model are
substantially more energetic than the ground states. For instance,
in $d=3$ relaxation gives a normal distribution of states of mean
$E_{\rm relax}/N \approx -1.4$ and a deviation of approximately
$0.1$, whereas the ground states found at $L=6$ are centered just
above $E_{\rm gs}/N\approx-1.7$, with a much narrower deviation and
close to the thermodynamic value of $-1.700(1)$.~\cite{pal:96a}

Finally in this section, let us return to Fig.~\ref{fig:TextraPh0_083} now 
concentrating on the gross $T$-dependence. 
This figure demonstrates another robust feature, the linearity of $P(h=0,T)$
with $T$ up to the mean-field transition temperature  $T_{MF}=1$;
this result previously observed  in the SK model
\cite{thomsen:86} is seen from the main figure to hold well also
for the KAS model with $\sigma=0.83$,
even though the actual transition temperature in this case is much smaller 
than the mean-field
temperature (indeed it is closer to $T_c=0.45$ \cite{katzgraber:05c}). As shown in the inset, a
similar behaviour is found for other values of $\sigma$ (up to the order of 
the peak in Fig.~\ref{fig:sigma_max}), the higher of which have no 
finite-temperature phase transition. Thus 
again the general shape of $P(h=0, T)- P(h=0,T=0)$ is rather robust against
whether the system is one which orders or not, has RSB or not. 

\section{Conclusions} 
\label{sec:conclusions}

We have presented results for the local field distributions of
different spin-glass models in different space dimensions. These
include both cases where there is generally believed to be a
finite-temperature phase transition and others where no finite
temperature transition occurs. They also include cases where the
subtleties of replica symmetry breaking are believed to operate and
others where they do not occur or are in question.    Our results show
that the distributions of local field are qualitatively very similar
for all models with Gaussian-distributed interactions; the data for
$P({\mid}h{\mid}, T=0)$ decay
exponentially for large fields $|h|$ and show a linear behaviour
for $|h|$ close to zero.  The distributions only show differences
in behaviour near $T = 0$ and $h = 0$ where for the infinite-range
SK model $P(h = 0, T = 0,N) \sim  N^{-1/2} \rightarrow 0$ as $N
\to \infty$, while all other finite-dimensional Edwards-Anderson
spin-glass models studied with space dimensions $d > 1$ $P(h =
0,N)$ seems to tend to a constant in the thermodynamic limit. For
finite $d$ the thermodynamic value of $P(0)$ scales as $d^{-1/2}$
for low $d$ with a coefficient very close to that for the SK model,
taking $2d=N-1$ to match coordination number for spins in EA and SK,
see Fig.~\ref{fig:extraPh0d}. This suggests that $d^{-1/2}$-scaling
applies for all $d$ \cite{comment:markus}.

These observations are mirrored by simulations of a one-dimensional
Ising model with random power-law interactions: for the regime of the
power-law exponent $\sigma$ which correspond to an infinite-ranged
system, $P(h = 0)$ decays with an inverse power of the system size,
whereas for all other universality classes the distributions tend to a
constant in the thermodynamic limit, rising from zero in the limits of
both SK $\sigma=0$ and the unfrustrated nearest-neighbour
$\sigma=\infty$, with a maximum for $\sigma$ near 2. Furthermore, for
all $\sigma$ the $P(h=0,T)$ are all close to the same linear-$T$
behaviour for $T<T_{MF}=1$.

Qualitatively similar, but quantitatively different, small-$h$ $P(h)$
behaviour is also found for systems that are quenched from random
starts (to states that are not thermodynamically equilibrated).

Our study has been concerned with the case of Gaussian exchange
distribution. For problems with discrete distributions, such as for
$J_{ij}$ randomly $\pm J$ one expects $P(h=0, T=0)$ to be nonzero
for finite-range systems \cite{thomsen:86}. It is, however,
surprising how small the observed $P(h=0,T=0)$ are for Gaussian
exchange-disorder without being zero. This effectively counsels
against associating small $P(0)$ and simple small-$h$ slopes with
the subtleties of RSB or finite-temperature spin glass transitions
\cite{comment:cg}.

Finally, we should caution that the simulations were performed for
finite system sizes and although corrections to scaling seem to be
very small, a change in behaviour at larger system sizes cannot be
ruled out completely.

\section*{Acknowledgments}

We would like to thank H.~Horner, M.~M\"uller, and G.~T.~Zimanyi
for helpful comments and suggestions.  S.B.~acknowledges support
from the Division of Materials Research at the National Science
Foundation under grant No.~0312510 and from the Emory University
Research Council.  H.G.K.~acknowledges support from the Swiss
National Science Foundation under grant No.~PP002-114713.  Part of
the simulations were performed on the Hreidar and Gonzales clusters
at ETH Z\"urich.  D.S.~acknowledges support from the UK Engineering
and Physical Sciences Research Council under grant No.~D050952.

\section*{References}
\bibliographystyle{spphys}
\bibliography{refs,comments}

\begin{thebibliography}{10}
\bibitem{comment:bound}
Actually, in Ref.~\cite{thouless:77} the authors give a bound on the
  coefficient but also express the belief that this bound should be saturated.
\bibitem{comment:ab}
It has been shown \cite{billoire:07} that system sizes of $N \gtrsim 200$ are
  needed to probe the thermodynamic limit of the KAS and SK models. The local
  field distributions $P(h)$ show small corrections to scaling and thus the
  system sizes studied in this work should suffice.
\bibitem{comment:cg}
A corollary is to anticipate in disordered semiconductors similar lack of
  connection between a ``Coulomb gap'' in the density of states centered at the
  Fermi level \cite{efros:75} and the existence or otherwise of a glassy state
  \cite{mueller:04,mueller:07}.
\end{thebibliography}

\end{document}